\documentclass[aps,10pt,twoco lumn,amsmath,amssymb,prl,superscriptaddress]{revtex4-1}
\usepackage{graphicx}
\usepackage{dcolumn}
\usepackage{bm}

\setcitestyle{super}

\begin{document}

\title{Nonlinear transport and noise thermometry in quasi-classical ballistic point contacts}
\author{E.S.~Tikhonov}
\affiliation{Institute of Solid State Physics, Russian Academy of
Sciences, 142432 Chernogolovka, Russian Federation}
\affiliation{Moscow Institute of Physics and Technology, Dolgoprudny, 141700 Russian Federation}
\author{M.Yu.~Melnikov}
\affiliation{Institute of Solid State Physics, Russian Academy of
Sciences, 142432 Chernogolovka, Russian Federation}
\author{D.V.~Shovkun}
\affiliation{Institute of Solid State Physics, Russian Academy of
Sciences, 142432 Chernogolovka, Russian Federation}
\affiliation{Moscow Institute of Physics and Technology, Dolgoprudny, 141700 Russian Federation}
\author{L. Sorba}
\affiliation{NEST, Istituto Nanoscienze-CNR and Scuola Normale Superiore, Piazza San Silvestro 12, I-56127 Pisa, Italy}
\author{G.~Biasiol}
\affiliation{CNR-IOM, Laboratorio TASC, Area Science Park, I-34149 Trieste, Italy }
\author{V.S.~Khrapai}
\affiliation{Institute of Solid State Physics, Russian Academy of
Sciences, 142432 Chernogolovka, Russian Federation}
\affiliation{Moscow Institute of Physics and Technology, Dolgoprudny, 141700 Russian Federation}

\begin{abstract}

We study nonlinear transport  and non-equilibrium current noise in quasi-classical point contacts (PCs) defined in a low-density high-quality two-dimensional electron system in GaAs. At not too high bias voltages $V$ across the PC the noise temperature is determined by a Joule heat power and almost independent on the PC resistance that can be associated with a self-heating of the electronic system. This commonly accepted scenario breaks down at increasing $V$, where we observe extra noise accompanied by a strong decrease of the PC's differential resistance. The spectral density of the extra noise is roughly proportional to the nonlinear current contribution in the PC $\delta S\approx2F^*|e\delta I|\sim V^2$ with the effective Fano factor $F^*<1$, indicating that a random scattering process is involved. A small perpendicular magnetic field is found to suppress both $\delta I$ and $\delta S$. Our observations are consistent with a concept of a drag-like mechanism of the nonlinear transport mediated by electron-electron scattering in the leads of quasi-classical PCs.  

\end{abstract}

\maketitle

%

The Landauer's approach~\cite{Landauer} accounts for elastic scattering of the charge carriers (electrons) off the potential inhomogeneities and is applicable in quasi one-dimensional conductors shorter than the phase-coherence length. Within this framework the temperature ($T$) and bias ($V$) dependencies of the conductance are caused solely by the averaging of the energy-dependent scattering matrices~\cite{vanHouten_beenakker}. Being a random process the resulting reflection (backscattering) of the electrons at the conductor produces non-equilibrium fluctuations of the electric current (shot noise)~\cite{blanter}. Qualitatively, increasing  backscattering reduces the current and increases the relative current fluctuation. The same holds true for a backscattering owing to the electron-phonon interaction in classical metallic point contacts (PCs)~\cite{kulik1984}. 

A different concept was recently proposed in Refs.~\cite{aivazyan,NagaevKost,nagaev_JETPL2011} for inelastic electron-electron scattering  ($e$-$e$ scattering) nearby a quasi-classical ballistic PC. Counter-intuitively, a reduction of the mean-free path owing to the $e$-$e$ scattering was predicted to give rise to an increase of the PC conductance~\cite{aivazyan}. In contrast to the backscattering scenario, this can be understood in terms of a drag-like interaction between the electrons of the incident and outgoing beams mediated by a non-equilibrium electronic distribution nearby the PC. In the linear response regime the PC conductance was predicted to increase linearly with $T$ in qualitative agreement with experiments~\cite{renard,melnikov2012}. In the nonlinear transport regime the same mechanism should give rise to the excess current contribution in the PC $|\delta I|\propto V^2$ and the excess shot noise with a spectral density of $2|e\delta I|$~\cite{nagaev_JETPL2011}. These hallmarks of the drag-like $e$-$e$ scattering mechanism we attempt to observe here. 

So far experimental observations of the leads-related noise were limited to $1/f$-like noise in classical metallic PCs with signatures of phonon emission~\cite{yanson1982} and trivial self-heating effect owing to Joule heat dissipation in semiconductor PCs~\cite{Kumar} and diffusive metallic nanowires~\cite{henny}. Recently~\cite{Leviton} a possibility to create minimal excitation states (called levitons)  in a two-dimensional quantum-PC was demonstrated using a shot-noise spectroscopy. Although one expects a decay of the levitons owing to the inelastic $e$-$e$ scattering, to which extent this could be traced via noise measurements remains unclear. Other available experiments were not optimized for refined studies of the $e$-$e$ scattering in the nonlinear transport regime. Backscattering off the depletion disk in scanning gate experiments~\cite{topinka2010} and phonon emission by extremely hot electrons in three-terminal devices~\cite{schinner} preclude observation of the extra current contribution predicted in Ref.~\cite{nagaev_JETPL2011}. 


Here, we report an experimental study of the nonlinear transport and noise in quasi-classical PCs in a two-dimensional electron system (2DES) in GaAs. Low carrier density, long elastic mean-free path and suppressed partition noise~\cite{blanter} make our samples well suited for 
this experiment. On top of the self-heating effect we observe extra noise contribution with the spectral density $\delta S$ accompanied by an extra nonlinear current in the PC $|\delta I|\sim V^2$. The extra noise is sub-Poissonian  $\delta S\approx2F^*|e\delta I|$, $F^*<1$, as expected for a random scattering process. Both $\delta I$ and $\delta S$ are suppressed in a small magnetic field perpendicular to the 2DES. In our opinion, these results elucidate the relevance of the drag-like $e$-$e$ scattering processes for the nonlinear transport in quasi-classical PCs and demonstrate a possibility for their detection via noise thermometry.

Our samples are made from two nominally identical (001) GaAs/AlGaAs heterostructures with 2DES buried 200\,nm below the surface. The electron density is about $0.9\cdot10^{11}\,{\rm cm^{-2}}$ and the mobility is $\approx4\cdot10^{6}\,{\rm cm^{2}/Vs}$ at $T=4.2$\,K, corresponding to elastic mean free path of $\approx$20\,$\mu$m. The split gate PCs are obtained with a standard e-beam lithography. We studied three samples with lithographic widths of the PC constriction 260\,nm and 600\,nm. Two (narrower) constrictions I and III had a T-shape geometry similar to that used in Ref.~\cite{melnikov2012}, while a (wider) constriction II had a standard symmetric shape. The linear response PC resistance $R_0$ is controlled by means of gate voltages applied to metallic split gates defining the constriction. The experiments were performed in a $^3$He/$^4$He dilution refrigerator (samples I, III) and in a $^3$He insert (sample II) at temperatures $T\approx0.1$~K and $T\approx0.5$~K, respectively, measured by Johnson-Nyquist noise thermometry. For shot noise studies we measured voltage fluctuations on a load resistor (1~k$\Omega$ for sample I and 3.3~k$\Omega$ for sample II) within the frequency range  10-20~MHz. In addition,  a tank circuit with a resonant frequency of about 13~MHz was used for sample II. In both cases the noise setup was calibrated by measuring the equilibrium Johnson-Nyquist noise of the device in parallel with the load resistor as a function of gate voltage at two different bath temperatures~\cite{suppl}. For samples I and II the nonlinear $I$-$V$ characteristics were obtained with a standard two terminal dc measurement. Sample III was used only for transport measurements in a four terminal scheme with a small ac modulation. The ohmic contacts to the 2DES were obtained via annealing of AuGe/Ni/AuGe and placed at distances about 1~mm away from the PC. The series resistance of each of the two 2DES regions connecting the PCs to the ohmic contacts is about $R_{\rm 2DES}\approx$\,180~$\Omega$ for sample I  and $R_{\rm 2DES}\approx$\,50~$\Omega$ for sample II. 

%
\begin{figure}[t]
 \begin{center}
  \includegraphics[width=\columnwidth]{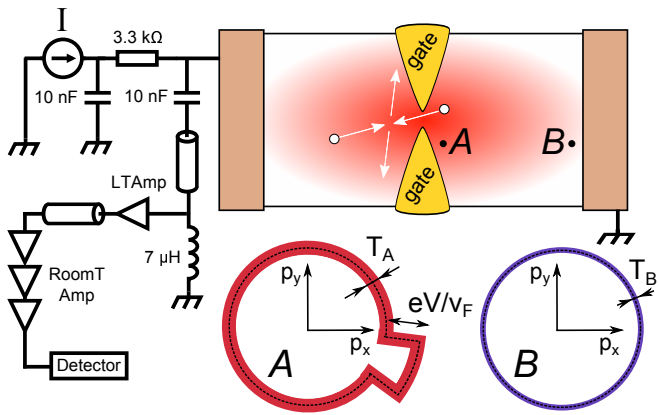}
   \end{center}
  \caption{Experimental layout. The PC constriction is formed with the help of split gates on the surface of the crystal. Current $I$ flows through the sample and drives the electronic distribution out of equilibrium. On top of the trivial self-heating of the 2DES, sketched by a color gradient, we are looking for a contribution of $e$-$e$ scattering of counter-propagating beams to current and noise  (white circles mimic scattering electrons). The momentum space distributions at points A and B are sketched in the lower part of the figure. In the former case, the distribution is anisotrpic with a bump of size $|eV|/v_F$ and a higher local temperature $T_A$, while in the latter case the distribution is locally equilibrium with a lower temperature $T_B$. The noise measurement scheme used for sample II is given on the left hand side.}\label{fig1}
\end{figure}

A layout of our experiment is shown in fig.~\ref{fig1}. Current flow through the split gate defined PC constriction drives the electronic system out of equilibrium. Further away from the PC the electronic distribution is of equilibrium Fermi-Dirac type characterized by a local temperature (see a sketch of a momentum space in point B). The local temperature reaches its maximum nearby the PC, as required by a balance of Joule heating and thermal conductivity (sketched by a color gradient in fig.~\ref{fig1}) --- a so called self-heating effect~\cite{Kumar}. In addition, next to the orifice, at distances smaller than the mean-free path, the electrons originating from different leads are not thermalized. Hence the electronic distribution  is anisotropic and characterized by a bump (in point A) or a dent of size $eV/v_F$ in momentum space~\cite{vanHouten_beenakker}, where $v_F$ is the Fermi velocity in the 2DES. In what follows we demonstrate that $e$-$e$ scattering of counter-propagating beams in the vicinity of the PC orifice predicted in Ref.~\cite{nagaev_JETPL2011} is responsible for the extra noise on top of the trivial self-heating.

\begin{figure}[t]
 \begin{center}
  \includegraphics[width=\columnwidth]{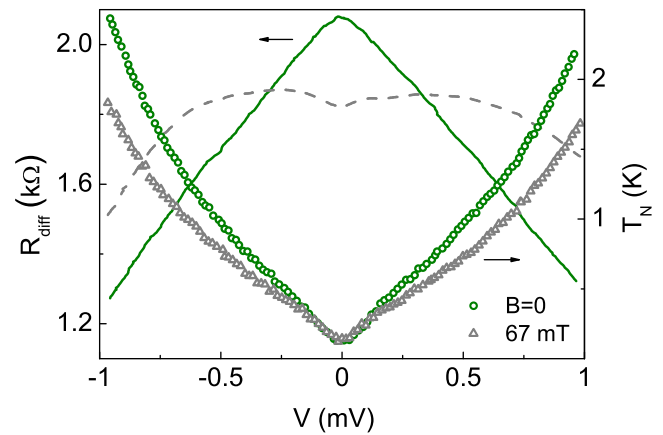}
   \end{center}
  \caption{Nonlinear transport regime in a quasiclassical PC. $R_{\rm diff}$ (lines, scale on the left) and $T_N$ (symbols, scale on the right) are plotted as a function of the bias voltage on the PC in sample I. The solid line/circles are taken in the absence of magnetic field, whereas the dashed line/triangles correspond to a perpendicular $B$-field of 67~mT. }\label{fig2}
\end{figure}

\begin{figure}[t]
 \begin{center}
  \includegraphics[width=\columnwidth]{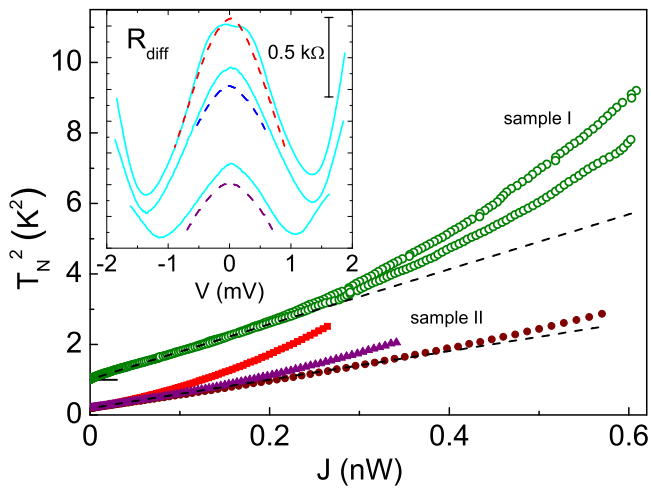}
   \end{center}
  \caption{Noise dependence on Joule heat power ($J$). Square of the noise temperature is plotted against $J$ for the two samples. Open symbols correspond to sample I and $R_0\approx\,2\,k\Omega$. Closed symbols correspond to sample II and  $R_0\approx\,0.5\,k\Omega$ (circles), $1.5\,k\Omega$ (triangles) and $3.4\,k\Omega$ (squares). In sample I the upper/lower trace corresponds to $V<0$/$V>0$, whereas in sample II the bias symmetry is preserved. Dashed lines are extrapolations of the linear dependencies $T_N^2-T_0^2\propto J$ at small $J$, where the equilibrium temperature is $T_0=0.1$\,K for sample I and $T_0=0.5$\,K and $R_0\approx\,1.5\,k\Omega$  for sample II. The data for sample I is shifted vertically for clarity, zero level indicated by a solid line. In this plot, the data for sample I were obtained in a separate run, compared to fig.~\ref{fig2}.	We believe that a difference of 15\% in $T_N$ between the datasets at $R_0\approx\,2\,k\Omega$ is an artifact owing to a long-term drift of the gain of the cryogenic amplifier, which was not figured out during the experiment. Inset: $R_{\rm diff}$ vs $V$ for sample II (dashed lines, $T_0=0.5$\,K) and $R_0\approx1.5\,k\Omega, 2\,k\Omega\,{\rm and}\,3.4\,{\rm k}\Omega$ from bottom to top and for sample III (solid lines, $T_0=0.1$\,K) and $R_0\approx1.2\,k\Omega, 2.3\,k\Omega\,{\rm and}\,3.3\,{\rm k}\Omega$ from bottom to top. The data is vertically shifted for clarity.}\label{fig3}
\end{figure}

Fig.~\ref{fig2} demonstrates typical experimental data for the differential resistance $R_{\rm diff}= dV/dI$ and current noise measured simultaneously in the nonlinear transport regime (sample I). Here we express the noise spectral density $S_I$ in terms of the noise temperature defined as $T_N=S_IR_{\rm diff}/4k_B$, where $k_B$ is the Boltzmann constant. At increasing dc bias voltage $|V|$ across the PC we observe that $R_{\rm diff}$ decreases by almost a factor of 2, roughly linear in $V$ (solid line). This observation is qualitatively analogous to the $T$-dependence of the linear response PC resistance~\cite{melnikov2012} and signifies a drag-like nonlinear contribution $|\delta I|\sim V^2$ in the $e$-$e$ scattering scenario~\cite{nagaev_JETPL2011}. Similar behavior is observed in all our samples, see the inset of fig.~\ref{fig3}. The decrease of $R_{\rm diff}$ in fig.~\ref{fig2} is accompanied by an increase of $T_N$ (right scale in fig.~\ref{fig2}). At not too high $V$ the bias dependence of the $T_N$ is close to linear, whereas an extra, nearly parabolic, contribution is well resolved at $|V|>0.5$~mV. 

We attribute the increase of $T_N$ at small $V$ to the self-heating of the 2DES~\cite{Kumar}. In this scenario one expects that $T_N$ equals the electronic temperature nearby the PC and is determined solely by a Joule heat power $J=I\cdot V\sim V^2$. To verify this conjecture we plot $T_N^2$ as a function of $J$ in fig.~\ref{fig3}. In both samples the data follow a linear dependence at small $J$, which is nearly independent of $R_0$ (shown only for sample II to save space). This is expected for a balance between the Joule heating and Wiedemann-Franz cooling~\cite{Kumar}: $T_N^2\approx T_0^2+R_{\rm 2DES}L^{-1}J$, where $T_0$ is the bath temperature, $R_{\rm 2DES}$ is the resistance of each of the two 2DES regions connecting the PC to the ohmic contacts and $L$ is the Lorenz number~\cite{remark_WF}. Quantitatively, the above formula underestimates $T_N$ in sample II by almost 40\%, which might be a result of additional heat resistance of the ohmic contacts, reduced thermal conductivity of the 2DES at higher $T_0$~\cite{lyahov} or a small residual partition noise. More important, however, is a lack of the universality observed in fig.~\ref{fig3} which becomes clearly visible at higher $J$ and $R_0$ and indicates extra noise on top of the self-heating. This is the same nearly parabolic in $V$ noise contribution we found in fig.~\ref{fig2}. Note that in sample I the extra noise depends on the sign of $V$, which correlates with the bias asymmetry of $R_{\rm diff}$ in fig.~\ref{fig2}, whereas in sample II the data is almost perfectly symmetric. We evaluate the extra noise temperature $\delta T_N$ as a difference between the measured $T_N$ and the self-heating contribution extrapolated linearly to high $J$, see dashed lines in fig.~\ref{fig3}. 


\begin{figure}[t]
 \begin{center}
  \includegraphics[width=\columnwidth]{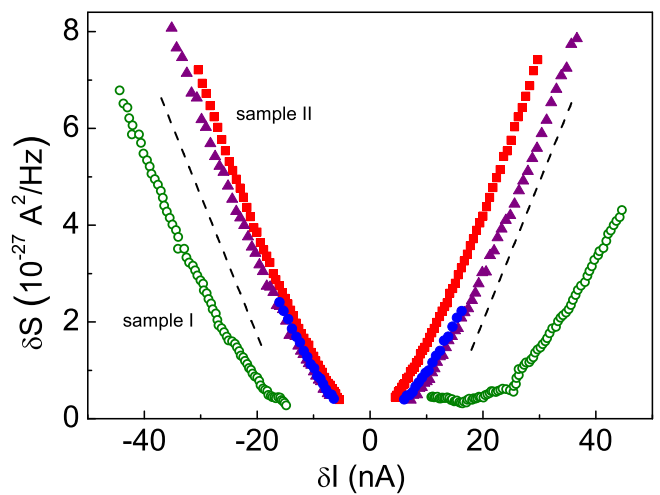}
   \end{center}
  \caption{Extra noise vs extra current. The spectral density $\delta S$ of the extra noise is plotted as a function of the nonlinear current contribution $\delta I$ both polarities of $V$. Open symbols correspond to sample I and $R_0\approx2k\Omega$ ($0.5\,{\rm mV}\leq|V|\leq1$\,mV). Closed symbols correspond to sample II and $R_0\approx\,1.5\,k\Omega$ (triangles, $0.4\,{\rm mV}\leq|V|\leq0.7$\,mV), $2\,k\Omega$ (circles, $0.4\,{\rm mV}\leq|V|\leq0.6$\,mV)  and $3.4\,k\Omega$ (squares, $0.5\,{\rm mV}\leq|V|\leq0.9$\,mV). Dashed lines are guides with a slope of $F^*=0.9$. For sample I, the data is obtained in the same run as in fig.\ref{fig2} and the scales on both axes are reduced by a factor of 2.}\label{fig4}
\end{figure}

In fig.~\ref{fig4} we plot a spectral density of the extra noise $\delta S=4k_B\delta T_N/R_{\rm diff}$ against a nonlinear current contribution through the PC $\delta I=I-V/R_0$. Our main observation here~\cite{remark_asymmetry} is that away from the origin $\delta S$ increases approximately linear with $\delta I$, that is both quantities follow nearly the same functional dependence on $V$. In the limit of high $\delta I$ the slope is consistent with the effective Fano factor of  $F^*\approx0.9$ (the slope of the dashed lines). Such a relation is expected for a random process with sub-Poissonian statistics, meaning that electrons carrying extra current are barely correlated. This is not possible for equilibrium Fermi distributions~\cite{Levitov_Lesovik} in the electron beams incident on the PC, suggesting that a random scattering process between the incident and outgoing electron beams in the 2DES nearby the PC is involved. Counter-intuitively, such extra scattering gives rise to the decrease of $R_{\rm diff}$, which perfectly fits in the $e$-$e$ scattering scenario of the nonlinear transport in quasi-classical PCs~\cite{nagaev_JETPL2011}. 

At $T=0$ uncorrelated scattering events between the carriers of the incident and outgoing beams result in a non-equilibrium electronic distribution and give rise to the Poissonian extra noise~\cite{nagaev_JETPL2011}. Sub-poissonian extra noise in fig.~\ref{fig4} does not necessarily imply correlations between the individual $e$-$e$ scattering events and can be explained by a finite $T$ effect associated with the self-heating. In our samples the self-heating results in $|eV|/k_BT<10$, a situation for which the numerical calculations predict extra shot noise with $F^*\approx0.5$ (see fig.~2 of the Ref.~\cite{nagaev_JETPL2011}). In spite of a qualitative agreement at higher $|\delta I|$, our data appreciably deviate from the expected linear dependence near the origin. On one hand, this deviation might stem from the uncertainty of our procedure to quantify the self-heating. It is this uncertainty that prevents us from analyzing the data at small $|V|$. On the other hand, there is an important difference between our experiment and theoretical assumptions. In Ref.~\cite{nagaev_JETPL2011}, the $e$-$e$ scattering is assumed weak compared to the elastic scattering, which determines the mean-free path and  cuts-off the logarithmic divergence of the collision integral. In contrast, we estimate~\cite{giulliani} the $e$-$e$ scattering length $l_{ee}\approx2\,\mu$m for electrons with the energy of $|eV|=$0.5~meV above the Fermi surface, i.e. an order of magnitude smaller than the elastic mean-free path in our devices. Still, in the regime $|eV|\gg k_BT_N$, the electronic distribution in the vicinity of the PC remains strongly anisotropic (sketch in fig.~\ref{fig1}) and the ideas of Ref.~\cite{nagaev_JETPL2011} are applicable qualitatively. Experimentally, this qualitative picture breaks down at even higher $|V|$, typically above 1~mV, where $l_{ee}$ becomes comparable to the width of the PC orifice. Here the bias dependence of $R_{\rm diff}$ changes sign (observed in all our samples and shown for sample III in the inset of fig.~\ref{fig3}), which can be explained by inelastic backscattering owing to multiple e-e scattering events in the vicinity of the PC. In support we observe that the upturn of $R_{\rm diff}$ shifts towards smaller $|V|$ with decreasing $R_0$ (i.e. increasing the width of the PC orifice). At the same time $T_N$ keeps increasing (not shown) apparently consistent with the suppression of thermal conductivity owing to the $e$-$e$ scattering~\cite{mishchenko}.

Finally, we study the effect of a small perpendicular magnetic field $B$ on the PC noise. The value of $B$ is chosen such that the cyclotron diameter (about 1.5 $\mu$m) is large compared to the width of the PC orifice and is much smaller than the elastic mean-free path. In this case the primary effect of $B$ is to bend the trajectories of scattering electrons and suppress the $e$-$e$ scattering contribution to conductance~\cite{NagaevKost,melnikov2012}. As shown in fig.~\ref{fig2}, in a magnetic field  (triangles) the overall $T_N$ decreases compared to the $B=0$ case (circles). At the same time, the decrease of $R_{\rm diff}$ with the bias voltage is suppressed in a magnetic field (dashed line in fig.~\ref{fig2}). These qualitative observations strongly support our interpretation of the nonlinear transport regime in terms of the drag-like $e$-$e$ scattering processes nearby the PC. We do not perform a quantitative account of the self-heating in a magnetic field, which is complicated in presence of a chiral heat flux along the edges of the sample~\cite{Prokudina_PRB}.

In summary, we studied nonlinear transport regime and noise in quasi-classical ballistic PCs. In addition to the trivial self-heating effect we observe extra noise with a nearly parabolic bias dependence. The extra contributions to the noise and current in the nonlinear transport regime are related via a sub-Poissonian value of the effective Fano factor, which is most likely a finite temperature effect. In addition, a small magnetic field perpendicular to the 2DES is found to suppress both. These observations provide evidence of the drag-like contribution of the $e$-$e$ scattering to the nonlinear transport and noise in ballistic PCs. 

We gratefully acknowledge discussions with V.T. Dolgopolov, K.E. Nagaev, and T.V. Krishtop. A financial support from the Russian Academy of Sciences, RFBR Grants No. 12-02-00573a and  No. 13-02-12127ofi, the Ministry of Education and Science of the Russian Federation Grant No. 14Y.26.31.0007 is acknowledged.


\begin{thebibliography}{50}
\bibitem{remark_WF} Here we neglected a Joule dissipation in the bulk 2DES, 
which results in corrections~\cite{Kumar} of order $R_{2DES}/R_0$.
\bibitem{remark_asymmetry} We do not discuss a pronounced bias asymmetry in sample I in fig.~\ref{fig4}, which is not reproduced in other sample and might be an artifact of our analysis.
\end{thebibliography}
\end{document}